
\magnification=1200 \vsize=23.5truecm \hsize=16truecm \baselineskip=0.7truecm
\parindent=1truecm \nopagenumbers \font\scap=cmcsc10 \hfuzz=1.0truecm
\def\yy2{\overline {\overline y}}
\def\trbc{\overline {\overline b}}
\def\trac{\overline {\overline a}}
\def\trcc{\overline {\overline c}}
\def\trdc{\overline {\overline d}}

\def\aa3{\overline {\overline {\overline a}}}
\def\bb3{\overline {\overline {\overline b}}}
\def\dd3{\overline {\overline {\overline d}}}
\def\yy3{\overline {\overline {\overline y}}}

\null \bigskip  \centerline{\bf THE GAMBIER MAPPING}
\vskip 2truecm
\centerline{\scap B. Grammaticos}
\centerline{\sl LPN, Universit\'e Paris VII}
\centerline{\sl Tour 24-14, 5${}^{\grave eme}$\'etage}
\centerline{\sl 75251 Paris, France}
\bigskip
\centerline{\scap A. Ramani}
\centerline{\sl CPT, Ecole Polytechnique}
\centerline{\sl CNRS, UPR 14}
\centerline{\sl 91128 Palaiseau, France}

\vskip 3truecm \noindent Abstract \smallskip
\noindent We propose a discrete form for an equation due to Gambier and
which belongs to
the class of the fifty second order equations that possess the Painlev\'e
property. In the
continuous case, the solutions of the Gambier equation is obtained through
a system of
Riccati equations. The same holds true in the discrete case also.  We use the
singularity confinement criterion in order to study the integrability of
this new
mapping.
\vfill\eject

\footline={\hfill\folio} \pageno=2

\noindent 1. {\scap Introduction. }
\smallskip
\noindent One century ago, Painlev\'e classified all second order of the form
$w''=f(w',w,z)$, where  $f$ is rational in $w'$, algebraic in $w$ and
analytic in $z$,
characterized by the absence of movable singularities [1]. This property, that
later came to be known as the Painlev\'e property, is considered today as
an indication
of integrability  [2]. Integrability was, of course, what Painlev\'e had in
mind
when he undertook his classification. The analysis presented by Painlev\'e,
however, was
not complete. Although he obtained most interesting results, discovering
the equations
that define the ``Painlev\'e transcendents'', his analysis  was plagued by
an oversight.
In particular he missed a whole class of equations the dominant part of
which reads:
$$x''={n-1\over n}{x'^2\over x}+\dots$$ As consequence he did not obtain
the general
forms of the transcendents P$_{\rm IV}$, P$_{\rm V}$ and P$_{\rm VI}$. The
gap in the
Painlev\'e classification was filled shortly afterwards by Gambier who
analyzed the
problem carefully [3] and presented the complete list of all 50 second order
equations that possess the Painlev\'e property.

Foremost among these new equations obtained by Gambier is equation XXVII in his
classification:
$$x''={n-1\over n}{x'^2\over x}+ axx'+bx'-{n-2\over n}{x'\over x}+{n
a^2\over (n+2)^2}x^3+
{n(a'-ab)\over n+2}x^2+fx-b-{1\over nx}\eqno(1.1)$$ This equation is
probably the most
complicated that Gambier had to study.  It is not one of the transcendental
equations,
neither is it integrable by quadratures. Rather, it belongs to the class of
equations
that can be integrated through linearization. The aim of the present paper
is to study
the discrete analog of this equation of Gambier. In previous publications
[4,5],
we have presented detailed studies of the discrete Painlev\'e equations,
while the case of
linearizable mappings was studied in [6]. However, in the latter, we have
limited ourselves to mappings that are linearizable through the discrete
equivalent of
the Cole-Hopf transformation. The Gambier equation needs a different
treatment. It is in
fact represented by two Riccati equations `in cascade'. In what follows, we
shall
present a brief review of the continuous case. This is motivated by the
fact that Ince's
book [7] (the standard reference on Painlev\'e equations) does not treat
the Gambier
equation in any detail. No indication as to how this equation can be
integrated is given,
and its particular cases are not even hinted at. Once the continuous
equation is properly
analyzed, its discretization follows in a most logical way.
\bigskip
\noindent 2. {\scap A brief recall of the continuous Gambier equation. }
\smallskip
\noindent The very essence of the Gambier equation is that it is obtained
as a cascade of
two Riccati equations [3]. One starts from a first Riccati:
$$y'=-y^2+by+c \eqno(2.1)$$ where $b,c$ are functions of the independent
variable $z$,
and couple $y$ to $x$ through a second Riccati:
$$x'=ax^2+nyx+\sigma \eqno(2.2)$$
 Here, $a$ is a function of $z$, $\sigma$ a constant that can be scaled to
1 unless it
happens to be 0, and $n$ is an integer. (As we shall see in what follows,
this last point
is a first requirement for the absence of critical singularities).
The quantities $a$ and $\sigma$ in (2.2) are related through a duality. Indeed,
replacing $x$ by $-1/x$ we find that the equation retains the same form
with $a$ and
$\sigma$ interchanged  (and $n\to -n$). Then if $a\neq 0$, the new $\sigma$
can be set to
1  (through  $x\to x/a$, and up to a translation of $y$: the new $a$
becomes $a\sigma$
instead of just $\sigma$).
In order to study the movable singularities  of the coupled Riccati system
we start from
the observation that from (2.1) the dominant behaviour of  $y$  can only be
$y\approx
{1/(z-z_0)}$. The next terms in the expansion of $y$ can be easily
obtained, and involve
the functions $b,$ $c$ and their derivatives.  In order to study the
structure of the
singularities of (2.2), we first remark that since the latter is a Riccati,
its movable
singularities are poles. However, (2.2) also has singularities that are due
to the
singular behaviour of the coefficients of the r.h.s of (2.2), namely $y$.
Now, the
locations of the singularities of the coefficients are `fixed' as far as
(2.2) is
concerned. However, from the point of view of the full system(2.1-2), these
singularities
are {\sl movable} and thus should be studied. The `fixed' character
reflects itself in
the fact -1 is {\sl not} a resonance. (The terms `resonance' is used here
following the
ARS terminology [8] and means the order, in the expansion, where a free
coefficient
enters. A resonance -1 is related to the arbitrariness of the location of the
singularity, and is thus absent when the location of the singularity is
determined
from the `outside' rather than by the initial conditions). Because of the
pole in $y$, $x$
has a singular expansion with a resonance different from -1 which may
introduce a
compatibility condition to be satisfied.

Before proceeding to the examination of the general case (2.2) let us
briefly consider
the case
$a=0$, whereupon (2.2) becomes linear. If, moreover, we take $\sigma$=0, we
find that the
behaviour of $x$ is obtained through $x'/x=ny$ and thus $x\approx
A(z-z_0)^n$. In this
case, since $n$ is integer, (2.2) always has the Painlev\'e property and
the functions $b$
and
$c$ are free. Next, we turn to the case of the general linear (2.2):
$x'=nyx+\sigma$ with
$\sigma\neq 0$. Let us examine a singularity of the form $$x\sim \lambda
(z-z_0)\eqno(2.3)$$ We find
$\lambda=\sigma/(1-n)$ unless $n=1$. (This last case is excluded as being of
non-Painlev\'e type. Indeed, one easily finds that for $\sigma=1$, the
singular expansion
has a logarithmic branching point of the form $x\sim  (z-z_0)\log(z-z_0)$.)
Thus,
excluding the case
$n=1$, we study just the singular behaviour (2.3). The resonance associated
to the
singularity (2.3) turns out to be exactly $n-1$, which explains the
requirement that $n$
be an integer. For every value of $n>1$ we expect a resonance condition
(with complexity
increasing with $n$).  For any $n$, a sufficient condition is $\sigma=0$
(which, as we
have seen above,  is also necessary for $n=1$) and we have a further
possibility which,
for the first few orders of $n$, writes:

$$n=2\quad\quad b=0 $$
$$n=3\quad\quad 2b^2-b'-c=0 \eqno(2.4)$$
$$n=4\quad\quad 6b^3-7bb'+b''-8bc+2c'=0 $$  For $n<0$ there is no further
constraint. In
fact the sufficient condition is $a=0$ which we have assumed in this paragraph.

We now turn to the case of the full Riccati (2.3) with $a\neq 0$. In this case,
thanks to the duality, we can assume  $\sigma=1$  (otherwise interchanging
$a$ and
$\sigma$ we are back to the previous case). Again, only the singularity due to
$y$ can lead to trouble. Rewriting (2.2) as
${x'/ x}=ax+ny+{\sigma/x}$ for $y=1/(z-z_0)+\dots$ we remark that unless
$n=\pm 1$  a
behaviour of the form $x\sim (z-z_0)^n$ is impossible when $a\sigma\neq 0$.
For $n=1$, a
logarithmic leading behaviour will be present for $\sigma\neq 0$, for
nonzero $a$ exactly
as in the case $a=0$. Conversely the condition $\sigma=0$ is sufficient for
the absence
of a critical singularity for $n=1$ even for nonvanishing $a$. Similarly,in
a dual way,
for $n=-1$ the necessary and sufficient condition for the absence of a critical
singularity is $a=0$, irrespective of the value of $\sigma$.

Next we assume $n\neq \pm 1$, in which case it suffices to study the
singularities
$x\approx
\lambda (z-z_0)$, $(\lambda=\sigma /(1-n))$   and $x\approx
\mu/(z-z_0)$, $(\mu=-(n+1)/a)$. The first singular behaviour ($x\approx
\lambda (z-z_0)$) has a resonance at $n-1$,  which is negative for $n<1$
and thus does
not introduce any further condition.  For $n>1$, the resonance condition
can be studied
at least for the first few values of
$n$. We find that $\sigma=0$ suffices for the resonance condition to be
satisfied even
for $a\neq 0$. However, if we demand  $\sigma \neq 0$, we find the further
possibilities:
$$n=2\quad\quad b=0 $$
$$n=3\quad\quad 2b^2-b'-c+a\sigma/2 =0\eqno(2.5)$$
$$n=4\quad\quad  18b^3-21bb'+3b''-24bc+6c'+8ab\sigma-2a'\sigma=0$$

The second singular behaviour, $x\approx \mu/(z-z_0)$, has a resonance at
$-1-n$.  Thus
for $n>0$ this resonance is negative and does not introduce any further
condition,
while for  $n<0$ a compatibility condition must be satisfied.  We find that
for every
case $n<0$, $a=0$ is a sufficient condition for the absence of critical
singularity.
(This is not in the least astonishing given the duality of $a$ and
$\sigma$). On the
other hand if we demand
$a\neq 0$ then a different resonance condition is obtained, at each value of n.
For $n=-1$, whenever $a\neq 0$, a logarithmic singularity of the form
$(z-z_0)^{-1}\log
(z-z_0) $ appears irrespective of the value of $\sigma$. For $n<-1$, we find:
$$n=\!-2\quad  a'-ab=0 $$
$$n=\!-3\quad 4ab^2-2ab'-2ac+a^2\sigma-6a'b+2a''=0\quad\eqno(2.6)$$
$$n=\!-4\quad
a(18b^3\!-21bb'+3b''-24bc+6c'+8ab\sigma)+a'(12b'+12c-33b^2-6a\sigma)+18ba''\
!-3a'''\!=0$$

A remark is in order at this point. The analysis presented above was based
explicitely on
the assumption that (2.1) is a Riccati. However, if we take $n\to\infty$ in
(2.2)  and
rescale $y$ (and $c$) in an appropriate way we find that (2.1) becomes {\sl
linear}:
$y'=by+c$. Thus, while on the one hand $n\to\infty$ in (2.5) leads to a
resonance condition that should be implemented after an infinite
number of steps, on the other hand we do not have to consider it. Equation
(2.1)
being now linear, $y$ does not have {\sl any} singularity that could
interact with
(2.2). The equation for $x$ reads in that case:
$$x''={x'^2\over x} +\left(ax+b-{\sigma\over x}\right)x'+(a'-ab)x^2+cx-\sigma
b\eqno(2.7)$$ which is a form of the equation XIV in the
Painlev\'e-Gambier classification [3,7] (not the canonical one but an
equivalent, generic
one). We recall that the canonical form of the latter  is:
$$w''={w'^2\over w} +(qw+{r\over w})w'+q'w^2-r'\eqno(2.8)$$
Equation (2.7) can be integrated to a Riccati:
$$x'=ax^2+qx+\sigma\eqno(2.9)$$
where $q$ is given by $(q\phi)'=c\phi$ and $\phi$ is an auxiliary function
related to
$b$ through $b=-\phi'/\phi$.
\bigskip
\noindent 3. {\scap The discrete analog of the Gambier equation. }
\smallskip
\noindent  In order to derive the discrete analog of the Gambier equation
we start with
the observation that in the continuous case we have two Riccati's in
cascade. It is well
known [6] that the discrete equivalent of the Riccati equation is a
homographic mapping.
Guided by this fact we can propose the following form for the discrete
Gambier system:
$$\overline y={by+c\over y+1}\eqno(3.1)$$ coupled to
$$\overline x={xyd+\sigma\over 1-ax}\eqno(3.2)$$ where $\overline x\equiv
x_{m+1}$,
$\overline y\equiv y_{m+1}$, and the $x$, $y$ are understood as $x_m$,
$y_m$. Appropriate
scalings have been made in the two equations in order to bring them to the
forms
(3.1-2). All four $a$, $b$, $c$, $d$ are functions of the discrete variable
$m$ while
$\sigma$ is a constant that can be set to 1 unless it is zero. Let us
remark that
$d$ may not vanish identically, since in that case the mapping (3.2) would
decouple.
(From (3.1-2), it is straightforward to obtain a 3-point mapping for
$x$ alone, but the analysis is simpler if we deal with both
$y$ and
$x$).

The main tool for the investigation of the integrability of discrete
systems is the
singularity confinement criterion we introduced in [9]. It is based on a
conjecture that
the movable (i.e. initial condition dependent) singularities of integrable
mappings
disappear after some iteration steps.  It will be used here to investigate the
integrability of the coupled mapping (3.1-2).

A first remark before implementing the singularity confinement algorithm is
that the
singularities of a Riccati mapping are automatically confined. Indeed, if
we start from
$\overline w=(\alpha w+\beta)/(\gamma w+\delta)$ and assume that at some step
$w=-\delta/\gamma$, we find that $\overline w$ diverges  but $\overline
{\overline  w}$
and all subsequent $w$'s are finite. Thus, the intrinsic singularities of
(3.2) do not
play any role. However, the singularities due to $y$ (obtained from (3.1))
may cause
problems at the level of (3.2). Two
types of singularities may appear. Either $y$ diverges (related to the
singularity
pattern  for $y$: $\{\dots,-1,\infty,b,\dots\}$) or $y$ takes the value
$y=-a\sigma/d$.
In the latter case, we find $\overline x=\sigma$ irrespective of $x$, and
thus $x$ loses
one degree of freedom. On the other hand, once we enter a singularity there
is no way for
$x$ to leave it unless $y$ assumes again a singular value (after a certain
number of
steps). Thus the first requirement for confinement would be for $y$ to
become again
either $\infty$ or equal to $-a\sigma/d$. However,  if $y$ were to take the
value
$\infty$ again some steps after first taking it, it would take it
periodically and the
singularity would then also be periodic. This is contrary to the
requirement that the
singularity be movable: a periodic singularity (with fixed period) is
`fixed' in our
terminology. The same applies to the singularity $y=-a\sigma/d$. Thus the only
singularity pattern for $y$ that are acceptable are: either start with
$-a\sigma/d$ and reach
$\infty$ after $N$ steps, or start with $\infty$ and reach $-a\sigma/d$
after $N'$
steps.
(Clearly, these two singularity patterns are mutually exclusive: one cannot
combine them
since in that case $\infty$ would be followed by $\infty$ after $N+N'$
steps and thus the
singularity would again be periodic). So the first condition is that
$y$ take the value
$-a\sigma/d$
$N$ steps before (or $N'$ steps after) taking the value $\infty$. The
equivalent of this
requirement in the continuous case is that the resonance be integer. We see
here that
this condition becomes here more complicated. In fact, the first few read:
$$N=1\quad-a\sigma/d=-1$$
$$N=2\quad-a\sigma/d=-{c+1\over b+1}\eqno(3.3)$$
$$N=3\quad-a\sigma/d=-{(\overline b+1)c+\overline c+1\over (\overline
b+1)b+\overline
c+1}$$ In what follows, we shall label these cases for $N>0$ while the
cases where the
value $-a\sigma/d$ appears $N'$ steps after $\infty$, such as:
$$N'=1\quad-a\sigma/d=\underline b\eqno(3.4)$$
$$N'=2\quad-a\sigma/d={\underline b\,{\underline {\underline b}}+\underline
c\over
{\underline {\underline b}}+1}$$
will be labelled by $N\equiv -N'<0$.

While one can compute easily more such conditions, it is not possible to
give a general
expression, like the ``$n$ must be integer'' of the continuous case.

Once $y$ hits a second singularity there is a possibility for $x$ to
recover its lost
degree of freedom through some indeterminate form (like $0.\infty$ or
$0/0$). Thus when
$y$ becomes infinite ($N$ steps after assuming the value $-a\sigma/d$) we must
simultaneously have $x=0$. Similarly when $y$ assumes the value
$-a\sigma/d$ ($N'$ steps
after being infinite) $x$ must be equal to $1/a$ so as to lead to $0/0$ for
$\overline
x$. We can study the first few cases and obtain the condition for the
confinement of the
singularity. Let us start with $N<0$. For $N=-1$, we enter the singularity
through
$y=\infty$, and thus ${\overline x}=\infty$,
${\overline y}=b$. We ask next that $-{{\overline a}\sigma/{\overline
d}}={\overline
y}=b$ and moreover that this lead to an indeterminate form in the calculation
of
$\overline {\overline x}$. The condition in that case is just $a=0$ (and
this implies
$b=0$ also, as $d$ identically zero is not allowed).

For $N=-2$ we obtain ${\overline{\overline x}}=-{\overline d}b/{\overline
a}$ and
ask that
$${\overline{\overline y}}={{\overline b}b+{\overline c}\over
b+1}=-{{\overline {\overline
a}}\sigma\over {\overline {\overline d}}}\eqno(3.5)$$ The condition for
$\overline{\overline {\overline x}}$ to result from an indeterminate form is
$1-{\overline {\overline a}}\,{\overline {\overline x}}=0$ i.e. ${\overline
{\overline a}}
\,{\overline d}b+{\overline a}=0$.  This last condition is automatically
satisfied when
$a=0$.

In fact, it turns out that  for every $N<0$, $a=0$ is a sufficient condition
for
confinement. In this case, the second equation of the mapping becomes
linear. Obtaining
the conditions for the next higher $N$ is straightforward. For $N=-3$, for
instance, we
find:
$${\yy3}={b{\overline b}\,{\trbc}+b{\trcc}+{\trbc} {\overline c}+{\trcc}\over
b{\overline b}+b+{\overline c}+1}=-{\aa3\sigma\over\dd3}$$
leading to
$$ (b+1)({\overline a}\,\aa3\sigma -{\overline a}-\trac\,{\overline d}b)-
\aa3\,\trdc\,{\overline d}b(b{\overline b}+{\overline c})=0\eqno(3.6)$$
The second possibility is to enter the singularity through  $y=-a\sigma/d$,
i.e. the
$N>0$ cases. For $N=1$, we enter the singularity through  $y=-a\sigma/d$
which gives
${\overline x}=\sigma$ and we exit through
${\overline y}=\infty$, which implies $a\sigma/d=1$. In order to have an
indeterminate
form at the level of ${\overline {\overline x}}$ we must have ${\overline
x}\, {\overline
y}=0.\infty$ and thus
$\sigma=0$ a condition clearly incompatible with $a\sigma/d=1$. Thus (3.1-2) as
written cannot confine for $N=1$ . In fact, the confinement condition would
be $\sigma=0$
{\sl provided} the denominator of (3.1) is just $y$ instead of $y+1$ (which
is dual
to $N=-1$ with $a=b=0$). However, for $N>1$ we will find that it is
possible to recover
the lost degree of freedom. For $N=2$, we ask that
$\overline{\overline {\overline x}}$ be zero and it turns out that this is
possible
if either $\sigma=0$ or ${\overline  d}=1$. In fact, the condition $\sigma=0$
is
sufficient for all cases $N>1$ (provided, of course, that the condition for
$N$, i.e.
``$y$ becomes infinite after
$N$ steps'', holds) For $N=3$, we write the condition (based on the assumption
$\sigma\neq 0$):
$$ (b+1)(a\sigma +{\overline d}-1)-(1+c)d{\overline d}=0\eqno(3.7)$$
Conditions for the next higher $N$'s can easily be obtained.

A question that naturally comes to the mind is whether the continuous limit
of system
(3.1-2) is indeed the coupled-Riccati original system of Gambier. The
answer appears
trivially affirmative at first sight: in fact, system (3.1-2) was
constructed as a direct
discretization of Gambier's system. However, a subtle question remains. In
equation
(2.2) of Gambier,
$n$ appears explicitely:  what is its equivalent in the discrete case? On
the one hand,
it is always possible to present a formal limit of (3.1-2) where $n$ is
related to $d$;
in fact we have $d=n-1$. However, in the previous paragraph we have argued
that what plays
the role of the position of the resonance in the discrete case is a
complicated relation.
This condition alone is not sufficient in order to ensure a form like (2.2)
at the
continuous limit with $n$ appearing naturally. (In particular, it does not
prevent $d$
from being non constant, which at the continuous limit would have given a
$n$ function
of $z$). Only when one takes into account {\sl both} confinement conditions
does system
(3.1-2) turn out to be of the precise form (2.1-2) at the continuous limit
for the
corresponding $n$. Moreover the `continuous' $n$ is precisely the `discrete'
$N$
introduced by the confinement. (Note, for instance, that the second
confinement condition
for $N=2$ is, precisely, $d=1$ i.e. $n=2$).

The discussion presented above was based on the form (2.1) for the first
Riccati. However
one should also consider another case that cannot be
obtained as a limit of (2.1) with the scaling we chose, namely:
$${\overline y}=b+{c \over y} \eqno (3.8)$$
Once the denominator has been simplified a new scaling is possible and we
rewrite
(3.8) as:
$${\overline y}=b+{1 \over y} \eqno (3.9)$$
The analysis of the coupled system (3.9) and (2.2) proceeds along the same
lines as
before. One obtains simpler discrete forms of the Gambier equation with
only one free
function instead of two in the generic case. As we have already remarked,
this is
the only case that confines for $N=1$, the condition being still $\sigma=0$.

****************************************************************************

Basil Grammaticos
LPN, Univ. Paris 7                   Tel-Fax Jussieu (33-1) 44277979
Tour 24-14, 5e etage                 Tel Polytechnique (33-1) 69334094
75251 Paris, France                  Tel Polytechnique (33-1) 69333008

****************************************************************************

\bigskip
\noindent 4. {\scap Discrete systems related to the Gambier mapping. }
\smallskip
\noindent  In the previous section, we have examined the mappings
consisting in two
Riccati's in cascade. Moreover, the particular case where the second
Riccati becomes a
linear equation was included in our study through $a=0$ and, because of the
duality
of $a$ and $\sigma$, through $\sigma=0$ as well. Finally the case
where the first Riccati reduces to  ${\overline y}=b+c/y$ was briefly
considered
in section 3 since it does not alter qualitatively the conclusions reached
with the full
(3.1).
In this section, we are going to consider  two particular cases of the
Gambier mapping.
The first case corresponds to equation (3.1) becoming linear. We thus have
the mapping:
$${\overline y}=y+c \eqno (4.1)$$
$${\overline x}={xyd+\sigma \over 1-ax}\eqno (4.2)$$
(where by the appropriate scaling we have taken $b=1$).

Clearly, this mapping is a discrete form of the equation P$_{14}$
encountered in section
2. This can be readily seen when we consider its continuous
limit. Putting
$x=w/\epsilon$, $d=\epsilon q(z)$, $a=\epsilon^2p(z)$, $c=-s+\epsilon
r(z)/d$ where
$s=q'/q$, we obtain indeed, at the limit $\epsilon\to 0$:
$$w''={w'^2\over w} +\left(pw+s-{\sigma\over
w}\right)w'+(p'-sp)w^2+(r+s^2)qw-\sigma b\eqno(4.3)$$ i.e. exactly (2.7),
i.e. P$_{14}$,
 in noncanonical form.

We thus expect system (4.1-2) to be integrable without constraint. What does
singularity confinement tell us on this point? Before proceeding to the actual
calculation, let us try to find the answer by analogy. The case of a
linear+Riccati
system was obtained in the continuous case by the limit $n\to \infty$.
Since the number
$N$ of steps before confinement is related to the position $n$ of the
resonance  in the continuous case (in fact $n=N$) we expect
confinement to be obtained only after {\sl an infinite number of steps!}
This turns out
indeed to be the case. Given the structure of the mapping (4.1-2) the only
singularity
can be
$y=-a\sigma /d$ (where (4.2) loses one degree of freedom). However, the
lost degree of
freedom cannot be recovered in a finite number of steps since we need
$y=\infty$. Thus the
mapping has a very peculiar behaviour: it can only confine after
an infinite number of iterations steps. More on this subtle point will be
given in the
next section.

Let us now turn to a second particular case of the Gambier mapping. Suppose
now that the
second equation (3.2) of the mapping has a denominator equal to $-ax$ instead
of
$1-ax$. To avoid a trivial reduction we must here take $\sigma\neq 0$. Then
(after a
rescaling and a redefinition of the parameters) the system becomes:
$${\overline y}={by+c \over y+1} \eqno (4.4)$$
$${\overline x}={xyd-1 \over x}\eqno (4.5)$$
Here also we have a peculiar behaviour with respect to singularity
confinement. Clearly,
(4.5) loses one degree of freedom when $y=\infty$. Moreover, there is {\sl
no} possibility
to recover this lost degree of freedom for any value of $y$. Thus the
singularity never
confines. As we shall argue in the next section this is a genuinely nonconfined
singularity. Our argument can be further
strengthened by considering the continuous limit of (4.4-5). We put
$x=1+\epsilon w$,
$b=-1-\epsilon-\epsilon^2(q-1/2))$,
$c=b+\epsilon^2(-1/4+\epsilon(2q-1)/4+\epsilon^2p)$
and take just $d=-2+\epsilon$, the latter being clearly a restriction which
is sufficient in order to prove our point.  We find
$$w''=-2w'w-(w'+w^2)^2-2q(w'+w^2)-q+4p+{1 \over 4}\eqno (4.6)$$
This last equation is manifestly of non-Painlev\'e type (because of the
$(w'+w^2)^2$
term). Although (4.4-5) is not confining in general, there exists a
particular case where
one can recover confinement, namely when (4.4) becomes linear:
$${\overline y}={y+c } \eqno (4.7)$$
$${\overline x}={xyd-1 \over x}\eqno (4.8)$$
In this case, ther is no way for $y$ to become singular and thus the
mapping has only confined singularities. The continuous limit is
compatible with our conclusion. Indeed, putting $x=1+\epsilon w$, $d=d(z)$,
$c=-2\epsilon d'/d^2+\epsilon^3q/d$ we find at the continuous limit
$\epsilon \to 0$ the equation:
$$w''=-2w'w-{d'\over d}(w'+w^2)-q\eqno (4.9)$$
The latter is (a non canonical form of) equation P$_5$ in the
Painlev\'e-Gambier list
[3,7] and is obtained precisely as the derivative of a Riccati. (Note that
by P$_5$ we
mean here the fifth equation in the list and not the fifth Painlev\'e
transcendent, which
we usually denote by P$_{\rm V}$).

Thus, the confining subcases  of the Gambier mapping obtained in this
section are all
variants of the `discrete derivative' (with respect to one of the
parameters) of a
homographic mapping. This means that we can start with a mapping of the form
${\overline w}={\alpha w+\beta \over \gamma w+\delta}$,
solve for one of the parameters, take the discrete derivative of this
parameter and
assign an $m$-dependent value to it. Given the duality of the homographic
mapping under
the transformation $w\to 1/w$, only two different types of such mappings,
corresponding to
derivatives with respect to either $\alpha$ or $\beta$, exist.
\bigskip
\noindent 5. {\scap A discussion of the confinement property. }
\smallskip
\noindent In the preceding sections, we have repeatedly used the criterion
of the
confinement of singularities in order to characterize the integrable or not
character of the Gambier mapping under its various forms. Since the
behaviours of the
systems of section 4 were rather atypical as far as the confinement is
concerned,
we feel that some detailed discussion is necessary at this point.

The confinement property was initially introduced (and related to
integrability)
for rational mappings, the structure of which would lead to vanishing
denominators and
the appearence of divergences [9].
 However it was rapidly realized that this is not the only manifestation of
a singular
behaviour. Thus, in [6], we introduced the notion of singularity  so as to
include all
cases where the system loses one degree of freedom. Confinement of the
singularity in
this case means recovery of the lost degree of freedom through the
appearence of an
indeterminate form ($0/0$, $0.\infty$, etc.). One important condition (in
perfect
analogy to the continuous case) is that the singularity be `movable' i.e.
initial
condition dependent. Let us illustrate this last point in the case of the
Riccati
mapping:
$$w_{m+1}={\alpha w_m+\beta \over \gamma w_m+\delta}\eqno(5.1)$$
where $\alpha$, $\beta $, $\gamma $ and $\delta$ depend in general on $m$.
The mapping variable $w$ can very well become infinite for some $m$, but
this (movable)
singularity is immediately confined: infinity does not play a particular
role in a
homographic mapping. However, if at certain iteration the determinant
$\alpha\delta-\beta\gamma$ of the coefficient matrix vanishes the mapping
loses one
degree of freedom and moreover there is no way that it can recover it in
its subsequent
evolution. However this non-confined singularity is
perfectly acceptable since it is {\sl fixed} i.e. its location does not
depend on the
initial conditions.

Clearly, the Riccati mapping is too simple to exhibit any interesting
behaviour as far as the singularity confinement is concerned. As soon as we
enlarge our
scope and consider systems of two Riccati's in cascade (as in the case of
the Gambier
mapping) the behaviour becomes considerably richer. A first point must be
made about
periodic singularities. They are assimillated to fixed ones since they do
not have a
specific location, that could depend on the initial conditions, but rather
they extend
everywhere.

The main difficulty of the Gambier mapping comes from the fact that the
distinction
between fixed and movable singularities is somewhat artificial [10]. A movable
singularity in the first equation of the system enters in the second one
through a
coefficient and should thus be considered as `fixed', as far as the second
equation is concerned. (Let us recall again that the `movable'
singularities of a
homographic mapping (5.1) do not cause any difficulties). Still
its location  depends on the initial conditions of the system, considered
globally, and
must, thus, be treated as a movable singularity.

In section 4 we have encountered two pathological cases. In one of them, we
have argued
that confinement was possible (although after an infinite number of
steps) and thus that the equation was in principle confining. The second
equation was
rejected as non-confining since there was no possibility whatsoever to
confine. We are
aware that the distinction is subtle and lies on grounds that are not very
firm.  In the cases at hand, we were guided by the underlying structure of
the system in
order to make our claims on the confining or not character of the mapping.
However, if we are given a general 3-point mapping, there is a priori no easy,
algorithmic, way to decide whether the mapping confines after an infinite
number of steps, or not at all. We hope that
just as in the continuous case, the occurence of a resonance at $n=\infty$,
compatible with the Painlev\'e property, is an exceptional occurence, and
that the 3-point
mappings confining after an infinite number of iterations will be
restricted to those
of the Gambier family.

A useful guide as to the confining or not character   of the mappings (in
such tricky cases) is the study of all possible continuous limits. Our
conjecture is
that confining mappings have as continuous limits differential equations
that possess
the Painlev\'e property. Thus if for some mapping we are able to find some
non-Painlev\'e  continuous limit, this is a clear indication as to its
non-confining
character. However, the absence of such non-Painlev\'e limits does not
allow us to
draw clear conclusions: it may only mean that non-Painlev\'e continuous
limits do not
exist (or that we were not able to find them!) Still, the continuous limit
can be a
`last stand' argument for some mapping that is difficult to interpret.

Finally, let us point out  that all our discussion of Gambier's mapping and, in
fact, of Gambier's equation as well, is a discussion of the Painlev\'e property
rather than plain integrability. In a system such as (2.1-2) or (3.1-2) of
Riccati equations in cascade, we can always solve the first equation for $y$,
obtain $y(t)$ or $y_m$, and inject it into the second equation. The latter can
always be written as a linear second order differential equation for $x$, or a
linear 3-point mapping. So, in principle, the problem
can always be solved formally. The difficulty comes when one wishes to actually
compute $x$, in terms of contour integrals in the continuous case while $y$
has bad
analytic properties. The Painlev\'e property guarantees that one can perform
the
integration and obtain indeed the solution of $x(t)$ over the complex
$t$-plane.
In the discrete case the difficulty arises when one tries to compute $x_m$
in terms of
matrix products (the elements of which contain $y_m$). When $y_m$ has the wrong
properties some of these matrices are singular in such a way that degrees
of freedom are
irretrievably lost. Singularity confinement means that these lost degrees
of freedom are
recovered at some later stage.
\bigskip
\noindent 6. {\scap Conclusion. }
\smallskip
\noindent In this paper we have analysed a mapping consisting in two
discrete Riccati's in
cascade. This mapping is the discrete analogue to Gambier's equation
(number XXVII of the
Painlev\'e-Gambier list). The interesting feature of this discrete system
is the fact
that the singularitites can be confined after an arbitrary number of steps.
This is
quite unusual, since, in most cases encountered up to now, confinement was
obtained
after a small number of iteration steps. This new behaviour, encountered
for the first
time in the Gambier mapping has made possible the refinement of the whole
notion of
singularity confinement.

\bigskip {\scap References}.
\smallskip
\item{[1]} P. Painlev\'e, Acta Math. 25 (1902) 1.
\item{[2]} A. Ramani, B. Grammaticos and A. Bountis, Phys Rep 180 (1989) 159.
\item{[3]} B. Gambier, Acta Math. 33 (1910) 1.
\item{[4]} A. Ramani, B. Grammaticos and J. Hietarinta, Phys. Rev. Lett. 67
(1991) 1829.
\item{[5]} B. Grammaticos and A. Ramani, {\sl Discrete Painlev\'e
equations: derivation
and properties}, NATO ASI C413 (1993) 299.
\item{[6]} A. Ramani, B. Grammaticos and G. Karra, Physica A 181 (1992) 115.
\item{[7]} E.L. Ince, {\sl Ordinary Differential Equations}, Dover, London
(1956).
\item{[8]} M.J. Ablowitz, A. Ramani and H. Segur, Lett. Nuov. Cim. 23
(1978) 333.
\item{[9]} B. Grammaticos, A. Ramani and V.G. Papageorgiou, Phys. Rev.
Lett. 67 (1991)
1825.
\item{[10]}     A. Ra\~nada, A. Ramani, B. Dorizzi, B. Grammaticos, J.
Math. Phys. 26
(1985) 708.
\end